%% file: main.tex
\documentclass[10pt,twocolumn,letterpaper]{article}

\usepackage{wacv}
\usepackage{times}
\usepackage{epsfig}
\usepackage{graphicx}
\usepackage{amsmath}
\usepackage{amssymb}
\usepackage[ruled,vlined,linesnumbered]{algorithm2e}
\usepackage{booktabs}
\usepackage{caption}
\usepackage{subcaption}
\usepackage{pgfplots}
\usepackage{comment}
\usepackage{array, makecell} 
\usepackage[numbers,sort,compress]{natbib}
\usepackage{times,epsfig,graphicx,amsmath,amssymb}
\usepackage{times,graphicx,amsmath,amssymb,booktabs,tabulary,multirow,overpic,bbm,amssymb}

\def\wacvPaperID{285} 

\wacvfinalcopy 

\ifwacvfinal
\def\assignedStartPage{1} 
\fi

\ifwacvfinal
\usepackage[breaklinks=true,bookmarks=false]{hyperref}
\else
\usepackage[pagebackref=true,breaklinks=true,colorlinks,bookmarks=false]{hyperref}
\fi

\ifwacvfinal
\else
\fi

\usepackage{authblk}
\makeatletter
\renewcommand\AB@affilsepx{, \protect\Affilfont}
\makeatother

\input{math_commands.tex}

\begin{document}

\title{A Weakly Supervised Region-Based Active Learning Method for COVID-19 Segmentation in CT Images}

\author[1,2]{Issam Laradji}
\author[2]{Pau Rodriguez}
\author[2]{Frederic Branchaud-Charron}
\author[3,5]{Keegan Lensink}
\author[2]{Parmida Atighehchian}
\author[4,5]{William Parker}
\author[2]{David Vazquez}
\author[6]{Derek Nowrouzezahrai}

\affil[1]{issam.laradji@gmail.com}
\affil[2]{Element AI}
\affil[3]{Xtract AI}
\affil[4]{SapienML}
\affil[5]{University of British Columbia} 
\affil[6]{McGill University}

\maketitle

\begin{abstract}
One of the key challenges in the battle against the Coronavirus (COVID-19) pandemic is to detect and quantify the severity of the disease in a timely manner. 
Computed tomographies (CT) of the lungs are effective for assessing the state of the infection. Unfortunately, labeling CT scans can take a lot of time and effort, with up to 150 minutes per scan. We address this challenge introducing a scalable, fast, and accurate active learning system that accelerates the labeling of CT scan images. Conventionally, active learning methods require the labelers to annotate whole images with full supervision, but that can lead to wasted efforts as many of the annotations could be redundant. Thus, our system presents the annotator with unlabeled regions that promise high information content and low annotation cost. Further, the system allows annotators to label regions using point-level supervision, which is much cheaper to acquire than per-pixel annotations.  Our experiments on open-source COVID-19 datasets show that using an entropy-based method to rank unlabeled regions yields to significantly better results than random labeling of these regions. Also, we show that labeling small regions of images is more efficient than labeling whole images. Finally, we show that with only 7\% of the labeling effort required to label the whole training set gives us around 90\% of the performance obtained by training the model on the fully annotated training set. Code is available at: \url{https://github.com/IssamLaradji/covid19_active_learning}.
\end{abstract}

\section{Introduction}

\begin{figure}[t]
  \centering
  \includegraphics[width=\linewidth,trim={0.6cm 0.0cm 0.6cm 0},clip]{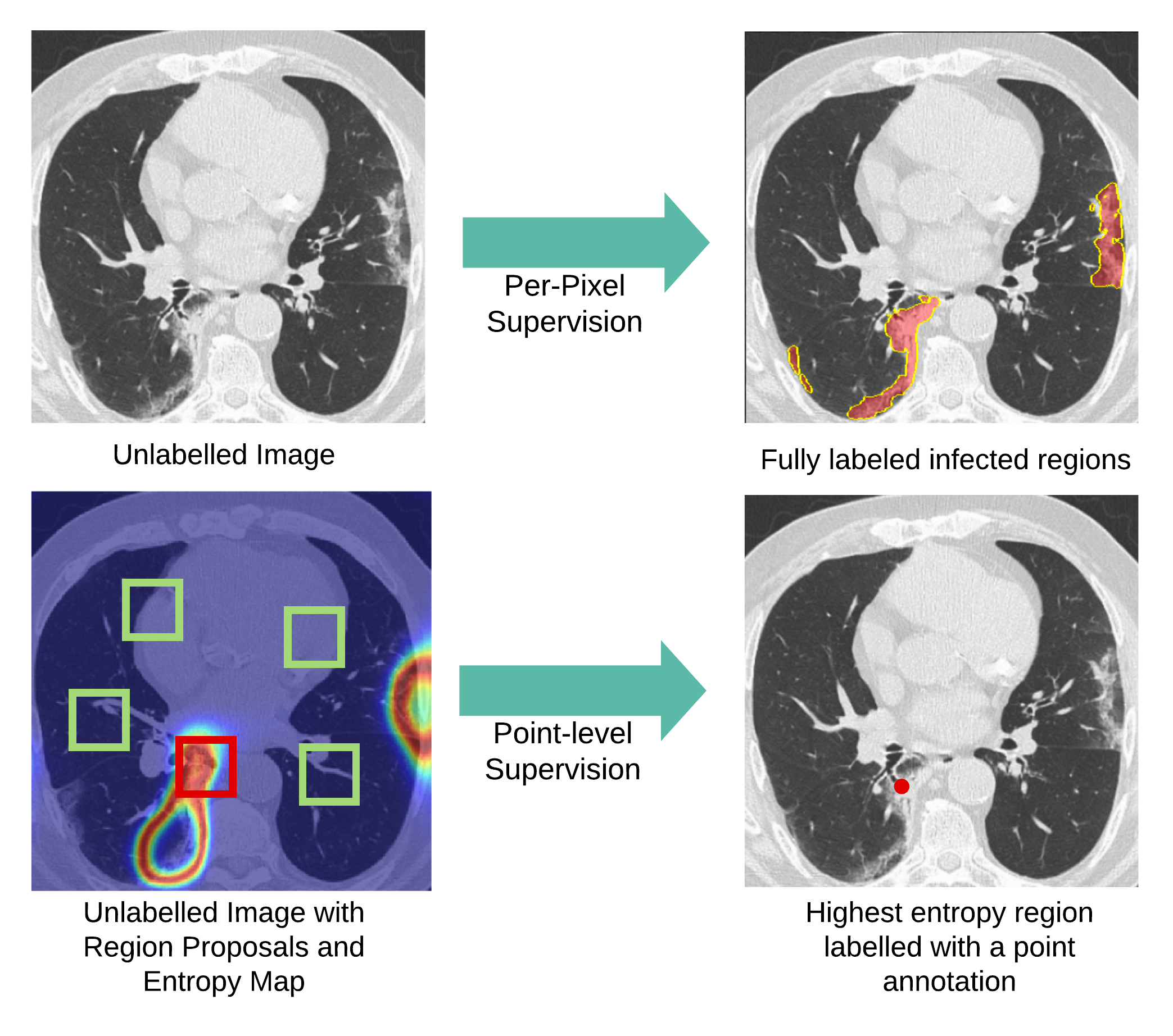}
  \caption{\textbf{Labeling Schemes.} (Top) Conventional per-pixel labeling  of the whole image. (Bottom) Our proposed region-based labeling scheme with point-level supervision. The region with the highest entropy (shown within the red rectangle) is labeled by clicking on a single pixel that is on top of an infected region.}
  \label{fig:model_uncertainty}
\end{figure}

The Severe Acute Respiratory Syndrome Coronavirus 2 (SARS-CoV-2) has rapidly spread into a pandemic and overwhelmed healthcare centers around the world.
While the disease (COVID-19) presents with a variety of symptoms, the build up of fluid in a patient's lungs has been most commonly associated with morbidity and mortality.
These affected regions, which are known as pulmonary opacification \cite{Hansell2008}, present as various patterns of attenuation on CT imaging and have been correlated with the severity of the COVID-19 infection \cite{Li2020, Wang2020}.
In severe cases, treatment of the disease requires intervention with essential equipment, which has lead to shortages around the world.
Accurate and accessible diagnostic methods are necessary to slow the spread of the virus, and efficient methods for prognosis and treatment are needed to ease the burden on healthcare centres in heavily affected regions. 

\begin{figure*}[t]
  \centering
  \includegraphics[width=0.65\linewidth,trim={0.6cm 0.8cm 0.6cm 0},clip]{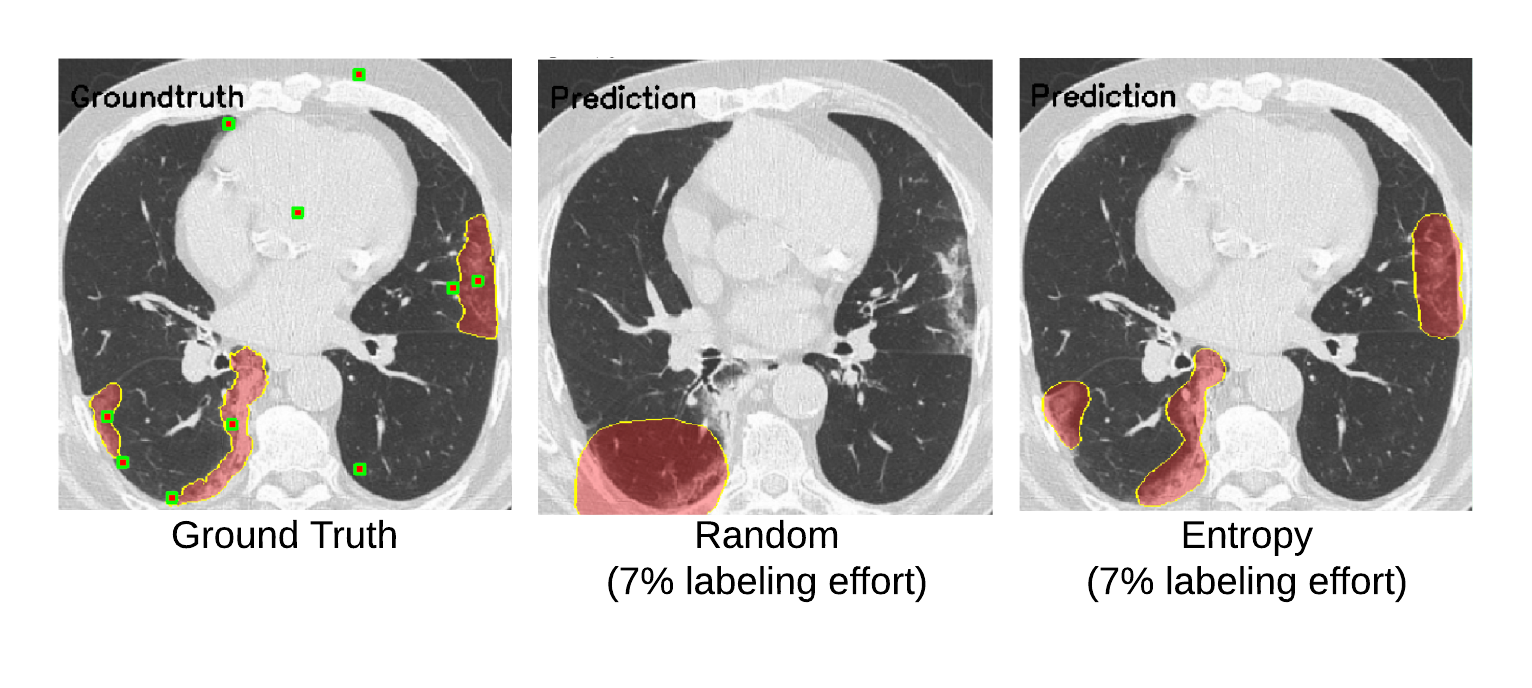}
  \caption{\textbf{Prediction comparison between fully supervised and region-based active learning system.} With only 7\% of the labeling effort (columns 2 and 3), segmented regions are close to the ground truth labels baseline (column 1). The points in column 1 represents example point-level annotations on infected regions and background. }
  \label{fig:comparisons}
\end{figure*}

RT-PCR (Reverse Transcription-Polymerase Chain Reaction) has emerged as the standard screening protocol for COVID-19, however it is time consuming and has a high false-negative rate \cite{zu2020coronavirus}. 
Recent work has shown that the analysis patterns of pulmonary opacification on chest CT scans provides a complementary screening protocol that achieves sensitive diagnosis \cite{ai2020correlation}. 
Additionally, recent work has shown that quantification of pulmonary opacification allows for the prognostication of patients, as the percentage of well-aerated-lung has been shown to be a predictive measure of intensive care unit (ICU) admission and death \cite{Colombi2020WellaeratedLO} .
In areas with concentrated COVID-19 infections, radiologists are burdened with the time consuming task of analyzing CT scans.
To this end, we investigate AI-based models for the segmentation of pulmonary opacification, thus significantly reducing the burden on healthcare centers and providing important information for the diagnosis and prognosis of COVID-19 patients.

Thus, we consider deep learning methods, which is a class of AI that has been successful in the medical imaging field for diagnosis, monitoring, and treatment of a variety of infections. Deep learning has already been applied to the medical image segmentation the brain~\cite{de2015deep, chen2018voxresnet} lung~\cite{harrison2017progressive}, and pancreas~\cite{roth2015deeporgan}. The goal is to assign a class label to each pixel in the images, which involves detecting unhealthy tissues or the areas of interest. The classical U-Net~\cite{ronneberger2015u} is one of the main deep learning segmentation methods that was shown to achieve promising performance in medical segmentation. Extensions to U-Net emerged to tackle medical segmentation using methods that are based on attention and multi-tasking~\cite{bui2019multi}. Overall, deep learning-based methods consistently outperform traditional methods in the medical image segmentation task. 

Recently, deep learning methods were used to help in the diagnosis of COVID-19 infections~\cite{li2020artificial, butt2020deep, yan2020covid, voulodimos2020deep}. These methods range from standard architectures to anomaly detection models designed to help radiologists analyze chest X-ray images. For CT images, segmenting COVID-19 infections was performed using location-attention oriented, 3D CT volume-based~\cite{zheng2020deep}, and edge detection based models~\cite{fan2020inf}. However, these methods do not consider model's feedback when labeling the training set, leading to possibly inefficient efforts as some training images might have redundant information.

According to \citet{ma2020towards}, it takes around 400 minutes to delineate one CT scan with 250 slices. It is important that only the scans that maximize the model's performance are labeled for cost efficiency. We address this challenge by introducing an active learning system combined with weak supervision. Active learning (AL) is a popular procedure to select the most informative images to label. The goal is to maximize the validation score with as few images labeled as possible. The information of an unlabeled image is often measured using entropy, which estimates the uncertainty of a model's output on that image. This approach has been beneficial for semantic segmentation~\cite{Mackowiak2018CEREALSC}. Similar to~\citet{Mackowiak2018CEREALSC, Casanova2020ReinforcedAL}, our active learning system only presents parts of the unlabeled image to the annotator for labeling (Figure~\ref{fig:model_uncertainty}). It was shown that it is easier for the annotator to label regions and allows the annotators to further focus their efforts on labeling the most informative image patches. 

These methods, however, require the annotator to label each region with per-pixel labels. This labeling scheme leads to two main challenges. First, per-pixel labels require a lot of effort. Second, under the active learning setup, it is difficult to calculate how much effort each region requires. Background regions require less effort to label than having to draw boundaries around infected regions. In \citet{Mackowiak2018CEREALSC, Casanova2020ReinforcedAL}, effort was measured based on the percentage of pixels labeled, which is not accurate.

For our active learning system, the annotator is allowed to label uninfected regions with the \textit{background} tag and regions with infections by placing a single click randomly on an infection. This scheme is also much faster to acquire than per-pixel labels, and we can accurately assume similar efforts between regions.

We evaluated our active learning framework on the publicly available CT Scan datasets.\footnote{Obtained from https://medicalsegmentation.com/covid19/} Our work follows the common AL setup where training is made of cycles, and in each cycle a set of images is selected for labeling~\citep{settles2009active}.  In each cycle the trained model computes an uncertainty map on the unlabeled regions first. Then, a set of unlabeled regions are sampled based on their uncertainty scores so that the more uncertain ones are labeled first. This procedure completes one cycle, and it is repeated until the annotation budget is reached. The intuition behind this method is that it allows the model to learn from low-effort highly informative regions to learn to perform good segmentation. 

We summarize our contributions and results on the COVID-19 benchmarks as follows:
\begin{enumerate}
\setlength\itemsep{0mm}
    \item We propose the first framework that combines region-based active learning with point-level supervision.
    \item For the COVID-19 datasets, we show that using an entropy-based method to rank unlabeled regions yields to significantly better results than random labeling of these regions when fixing the annotation budget.
    \item For the same datasets, we show that region-based active learning leads to better results compared to whole image labeling.
    \item We show the point-level supervision yields better performance with respect to budget compared to per-pixel annotation.
\end{enumerate}


\begin{figure*}[t]
  \centering
  \includegraphics[width=\linewidth,trim={0.6cm 0.4cm 0.4cm 0},clip]{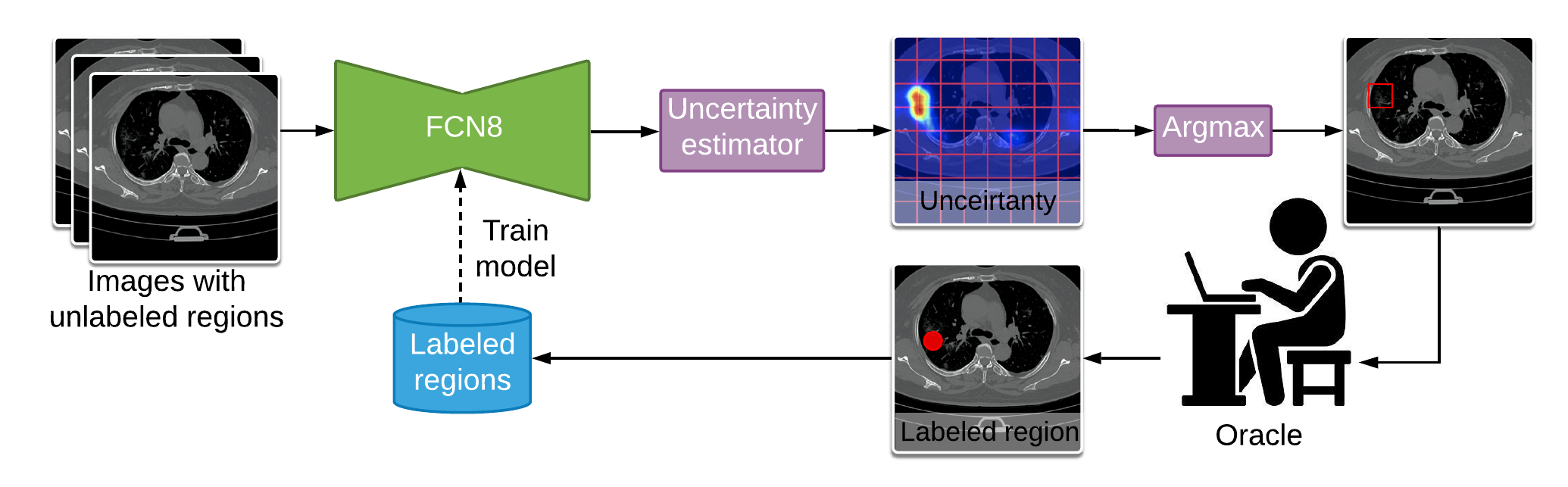}
  \caption{\textbf{Active Learning Setup.} \textbf{1.} We train the model on the labeled dataset. \textbf{2.} The trained model is used to estimate the uncertainty on all unlabeled images. \textbf{3.} The \textit{K} most uncertain regions are selected and labeled with point-level supervision. \textbf{4.} The newly labeled examples are added to the dataset for the next training cycle.}
  \label{fig:model_training}
\end{figure*}

\section{Related Work}
This work falls under the intersection between active learning, weakly supervised and semantic segmentation. We review the relevant work for each of these topics below.

\paragraph{Active learning} aims to maximize the performance on the test set with respect to the number of labeled examples. Different methods exist for selecting which data to be labeled from the unlabeled pool. These methods can be categorized into two categories. First, classical methods include query-by-committee~\cite{Dagan95committee, NIPS1992_622}, and ensemble disagreement~\cite{beluch2018power}. Secondly, Bayesian methods propose to sample from the posterior distribution before applying an heuristic on the set of predictions. Examples of the latter include \citet{maddox2019simple, gal2016dropout}. Moreover, different heuristics have been proposed to decide which samples to be labeled. These heuristics decide based on different strategies such as entropy~\cite{shannon_entropy}, maximizing the error reduction~\cite{roy2001toward}, or information theory~\cite{gal2017deep, houlsby2011bayesian}. The heuristics are often used to compute an uncertainty value for the whole image, whereas in this work we compute the entropy for different regions in the image to identify which object instances require per-pixel labels.

\paragraph{Active learning for semantic segmentation} is relatively less explored compared to classification, perhaps because of its challenging large-scale nature. Methods that work on this setup~\cite{dutt2016active} combine metrics that encourage the diversity and representativeness of labeled samples. Some rely on unsupervised superpixel-based over-segmentation~\cite{vezhnevets2012active, konyushkova2015introducing}. Others focus on foreground-background segmentation of biomedical images~\cite{melanoma-17, yang2017suggestive}. \citet{settles2008active, vijayanarasimhan2009s}, and ~\cite{Mackowiak2018CEREALSC} focus on cost-effective approaches, proposing manually-designed acquisition functions based on the cost of labeling images or regions of images.

Recent work on active learning with semantic segmentation relies on dividing the images into fixed-sized regions~\cite{Mackowiak2018CEREALSC, Casanova2020ReinforcedAL} and labeling the highest scoring ones with per-pixel labels. Unfortunately, these methods have two drawbacks. First, the size of the regions need to be predefined and the size can affect the performance widely. In many cases, it is more cost-effective to simply label a single object than a square region. Further, computing the labeling effort for a region is complicated. In \citet{Mackowiak2018CEREALSC, Casanova2020ReinforcedAL}, the labeling effort of these regions is assumed to be the same,  which is not always the case. Regions that have a single object class are much easier to label than regions with more than one object class. Second, per-pixel labels can be less cost-effective than weaker labels.

\paragraph{Active learning for medical segmentation} has received a lot of attention lately due to its potential in reducing the amount of human effort required to obtain a good training set. Acquiring medical datasets is difficult because it requires expert labelers (doctors) and long annotation time. As a result, there is a limited amount of labeled medical datasets compared to datasets from other domains. \citet{ceal2017gorriz} proposes to use the well-known CEAL method~\citep{wang2016cost} where uncertain examples are labeled by a human and confident examples are labeled by the model. They use U-Net~\citep{ronneberger2015u} with MC-Dropout~\citep{gal2016dropout} and estimate the uncertainty using the predictive variance. \citet{yang2017suggestive}  train an ensemble network and compute the similarity between features to estimate uncertainty. If the feature vectors are similar, the sample is easy and should not be annotated.

\paragraph{Weak supervision for semantic segmentation} can vastly reduce the required annotation cost for collecting a training set~\cite{Zhou2018PRM, khoreva2017simple, zhou2018weakly, laradji2019instance, laradji2019masks}. Collecting image-level and point-level labels for the PASCAL VOC dataset~\citep{everingham2010pascal} takes only $20.0$ and $22.1$ seconds per image, respectively~\citet{bearman2016s}. In comparison, acquiring full segmentation labels can take $239.0$ seconds per image on average. Other forms of weaker labels were explored as well, including bounding boxes~\cite{khoreva2017simple} and image-level annotation~\cite{Zhou2018PRM}. In this work, the labels are given as point-level annotations instead of the conventional per-pixel level labels.

\paragraph{Active learning with weak supervision} is a relatively new research area. To the best of our knowledge, it has only been investigated for the task of object detection~\cite{Chandra_2020, desai2019adaptive, Brust_2020}. \citet{Chandra_2020} and  \citet{desai2019adaptive} have proposed frameworks to use a combination of strong supervision and weak labels in the training process. Leveraging weak labels was shown to reduce the required annotation budget to attain good performance in the active learning setup for object detection. However, strong supervision was still required at the later stages of the training in order to achieve the optimal performance. 

\citet{Chandra_2020} have proposed a two-stage sampling method that is performed in every active learning cycle. In the first stage, images are sampled from the unlabeled data for which the oracle provides weak labels. In the second stage, images are sampled from the weakly labeled data for which the oracle provides strong labels. \citet{desai2019adaptive} have proposed an adaptive supervision method. By default the query is sampled from the unlabeled data and the oracle provides the weak labels. There are two conditions that define which level of supervision to use. For the first condition, if the prediction confidence is lower than a certain threshold, the images acquired in the current cycle are labeled with full supervision. The level of supervision for the second condition is based on the value of the loss. In our work, we are the first to combine region-based active learning with point-level supervision and apply it on the task of medical segmentation.

\section{Methodology}
\label{sec:methodology}

\begin{figure}[t]
  \centering
  \includegraphics[width=0.7\linewidth,clip]{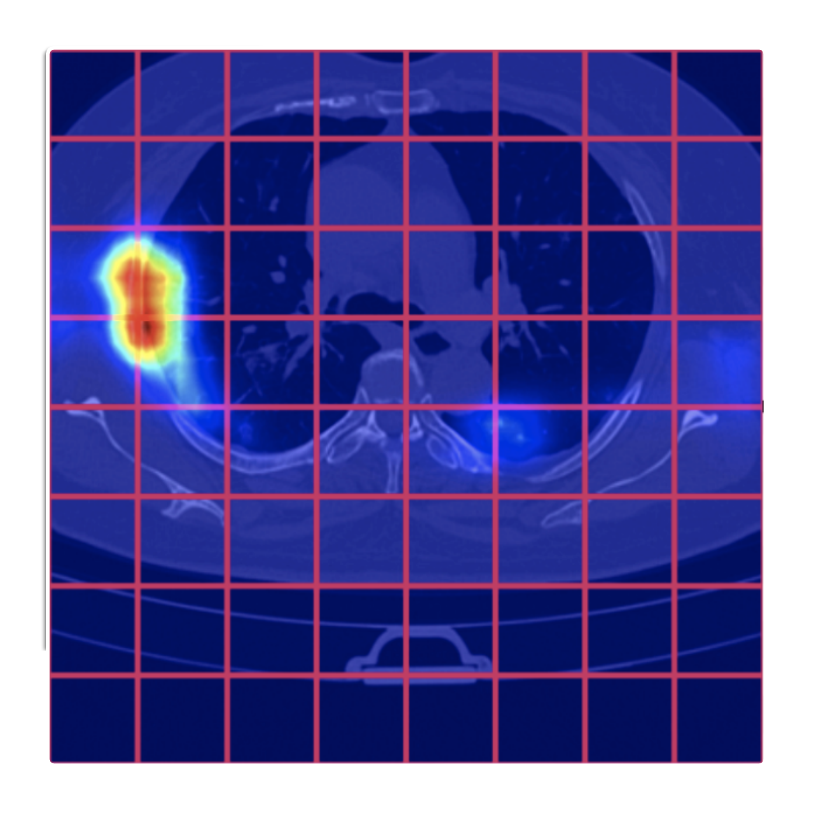}
  \caption{\textbf{Region-based Labeling.} This image is divided into 64 equally-sized non-overlapping rectangles, where each rectangle represents an unlabeled region. The region that has the highest per-pixel entropy mean (shown as heatmap) is selected for labeling.}
  \label{fig:grid}
\end{figure}

\paragraph{Setup.} As shown in Figure~\ref{fig:model_training}, we follow the common active learning setup where images are divided into labeled $\mathcal{X}_l$ and unlabeled $\mathcal{X}_u$ images. The process is divided into cycles. In each cycle the model is trained on $\mathcal{X}_l$ until convergence before the next batch of unlabeled examples are sampled from $\mathcal{X}_u$ for labeling. In the conventional active learning setup, the annotator is required to annotate every pixel in each sampled unlabeled image. This process might not be cost efficient as some regions could be very costly to label while having only little positive impact on the model's performance. 
Thus, we instead reformulate the problem by dividing each image into a grid of $K$ equal-sized rectangles as possible regions for labeling (see Figure~\ref{fig:grid}). In this case, the dataset is divided into $\mathcal{X}_l$, $\mathcal{X}_p$, and $\mathcal{X}_u$ where $\mathcal{X}_p$ is a set of partially labeled images. In each cycle, regions are sampled from the images $\mathcal{X}_p$, and $\mathcal{X}_u$ and are then  passed to a human oracle for labeling. These regions are selected based on either random or entropy-based heuristics. The latter heuristic allows the model to determine which regions it is mostly uncertain about, which can help improve its generalization performance when labeled. 

\paragraph{Labeling Scheme.} We consider two labeling methods: per-pixel and point-level. For per-pixel labels, the annotator is asked to label an object so that each of its pixels are annotated. Given $X$ as a set of $N$ training images with corresponding ground truth labels $Y$. $Y_i$ is a $W \times H$ matrix with the value of each entry corresponding to the class label.

When labelers are presented with an unlabeled region, they are only required to annotate that region with per-pixel labels (Figure~\ref{fig:model_uncertainty}). However, this type of annotation is costly because it requires the labeler to carefully draw a boundary around the object while dealing with occlusions and potential overlapping objects. According to the authors of the COCO dataset~\cite{lin2014microsoft}, it took around 22 worker hours for 1,000 segmentations. This annotation time implies a mean labeling effort of 79 seconds per object segmentation. Also, according to \citet{ma2020towards}, it takes around 400 minutes to delineate one CT scan with 250 slices. That is an average of 1.6 minutes per slice. While this labeling scheme is highly expressive, the information content it provides to the model might not be worth the labeling cost.

Thus, we also consider point-level labels. This labeling scheme allows the annotator to label a single point for each infected region. If the region has no infection, then the annotator is required to classify it as background. The ground truth mask $Y_i$ is a $W \times H$ matrix with entries 1 that indicate the locations of the infected regions, entries 0 that indicate background regions, and -1 that indicated unlabeled regions. The annotation cost for each region is similar.

\paragraph{Model Architecture.}
We use a segmentation network based on FCN8~\citep{long2015fully} with an ImageNet~\cite{imagenet_cvpr09} pretrained backbone. The network takes as input an image of size $W \times H$ and applies the forward function $f_\theta$, producing a $W \times H \times C$ per-pixel map where $C$ is the set of object classes of interest and $\theta$ are the network parameters. The output map is converted to a per-pixel probability matrix $S_i$ by applying the softmax function across these classes. These probabilities indicate the likelihood for each pixel of belonging to the infected region of a class $c \in C$. At test time, for each pixel, the class with the highest probability is selected.

\paragraph{Loss Function for Point-level Supervision.} 
We apply the standard cross-entropy function against the provided set of point-level annotations which represent the locations of the infected regions and the background pixels. The loss function is defined as follows,
\begin{equation} \label{eq:pointlevel}
\mathcal{L}_P(f_\theta,X_i,Y_i) = -\sum_{j \in \mathcal{I}_i}\log(f_\theta(X_i)_{jY_j})\;,
\end{equation}
where $f_\theta(X_i)_{jY_j}$ is the output corresponding to class $Y_j$ for pixel $j$, and $\mathcal{I}_i$ is the set of labeled pixels for image $X_i$.

\paragraph{Region-Selection methods.} In each cycle we select regions based on random, or entropy~\cite{gal2016dropout} heuristics. With the \textit{random} heuristic, regions are sampled randomly from $X_p$ and $X_u$. With the \textit{entropy} heuristic, the regions with the highest mean per-pixel entropy are selected. 

In order to obtain uncertainty measures based on the entropy of the semantic segmentation predictions, we add a dropout layer after fc6 and fc7 in the VGG16~\citep{simonyan2014very} architecture. Using MC-Dropout~\citep{gal2016dropout}, we acquire $I$ predictions drawn from the posterior distribution,
\begin{equation}
\label{eqn:bma}
    S_{ij} = f(x_i, \theta_j) \mid \theta_j \sim p(\theta \mid \mathcal{L}),
\end{equation}
where $S_ij$ is the predicted distribution per pixel for a model $f$ and an input $x_i$. This function allows us to select informative images for labeling. The intuition behind  MC-Dropout is that if the knowledge of the network about a visual pattern is precise, the predictions should not diverge if the image is evaluated several times by dropping weights randomly at each time.

We estimate the per pixel uncertainty  by computing the entropy of the mean estimator. Let $\hat{S_i}$ be the mean estimation over $I$ draws. We compute the uncertainty with: $U_i = \sum_c^C S_{ic}\log(S_{ic})$.

\paragraph{Model Training.} 
We start with an empty set of labeled images $X_l = \phi$. Then, we randomly sample an initial set of images and label them with per-pixel labels. Whenever we acquire a new labeled batch we train the model until convergence. In each cycle we compute the per-pixel entropy for for all unlabeled and partially labeled images. The score of each unlabeled region is the maximum pixel entropy within that region. The score of an image is the score of the region with maximum score. Images with unlabeled regions are then ranked based on their score. We then pick the K highest ranked images and select the highest scoring region from each image to labeled with point-level supervision. We terminate the training procedure after $T$ cycles.

\paragraph{Implementation Details}
Our methods use an Imagenet-pretrained VGG16~\cite{Simonyan2014VGG} FCN8 network~\cite{long2015fully}. Other Imagenet-pretrained architectures can be used as well, but we did not observe a difference in the results compared to other architectures such as UNet~\cite{ronneberger2015u} and PSPNet~\cite{zhaocvpr2019}. We ran the active learning procedure for 100 cycles. In the first cycle, 5 CT images were randomly sampled from the unlabeled pool and all their regions were labeled based on the required supervision level. Each image is divided into 64 equally-sized non-overlapping regions. In each cycle, 5 images are sampled from the unlabeled pool and for each of these images, a single region gets selected for labeling.  The maximum number of training epochs in a cycle is 40. The score is reported on the test set and it corresponds to the model that achieved the best score on the validation set. The models are trained with a batch size of 1 using the ADAM~\cite{kingma2014adam} optimizer with a learning rate of $10^{-4}$. To compute the uncertainty scores of an image, we perform  Monte-Carlo with samples following the procedure in~\citet{gal2016dropout}. The dropout rate was set to 0.5.


\begin{table}
\caption{Statistics of COVID-19 datasets.}
\label{tab:datasets}
\centering
\resizebox{\columnwidth}{!}{%
\begin{tabular}{lcccc}
\toprule
       Name &  \# Cases &  \# Slices  &  \# Slices with &  \# Infected \\
        &   &    &  Infections (\%) &  Regions \\
\midrule
 COVID-19-A &    9 &        829 &                       372 (44.9\%) &                1488 \\
 COVID-19-B &    20 &       3520 &                     1841 (52.3\%) &                5608 \\
\bottomrule
\end{tabular}
}
\end{table}

\section{Experimental Setup}

\subsection{Datasets}
\label{ssec:setup}
We evaluate our system on two open source datasets (COVID-19-A/B) whose statistics are shown in Table~\ref{tab:datasets}.


\paragraph{COVID-19-A~\cite{covid19a}} consists of 9 volumetric COVID-19 chest CTs in DICOM format containing a total of 829 axial slices. Images were first converted from Houndsfield units to unsigned 8-bit integers, then resized to $352 \times 352$ pixels and normalized using ImageNet dataset statistics~\cite{ILSVRC15}. Each axial CT slice was labeled for ground-glass, consolidation, and pleural effusion by a radiologist. We use two splits of the dataset: {\it separate} and {\it mixed}. In the separate split (COVID-19-A-Sep), the slices in the training, validation, and test set come from different scans. The goal is to evaluate how the model generalizes to new patients. In this setup, the first 5 scans are defined as training set, the 6th as validation, and the remaining as test. For the mixed split (COVID-19-A-Mixed), the slices in the training, validation, and test set come from the same scans. The idea is to evaluate if given few labelled slices from a scan the model can infer the masks for the remaining slices. For each scan, the first 45\% slices are defined as the training set, the next 5\% as the validation, and the remaining as test.

\paragraph{COVID-19-B~\cite{zenobo}} consists of 20 COVID-19 CT volumes. Lungs and areas of infection were labeled by two radiologists and verified by an experienced radiologist. Each three-dimensional CT volume was converted from Houndsfield units to unsigned 8-bit integers and normalized using ImageNet data statistics~\cite{ILSVRC15}. We also split the dataset into {\it separate} and {\it mixed} versions. For the separate split (COVID-19-B-Sep), we assign 15 scans to the training set, 1 to the validation set, and 4 to the test set. For the mixed split (COVID-19-B-Mixed), we separate the slices from each scan in the same manner as for COVID-19-A.


\begin{figure*}[t]
  \centering
  \includegraphics[width=0.8\linewidth]{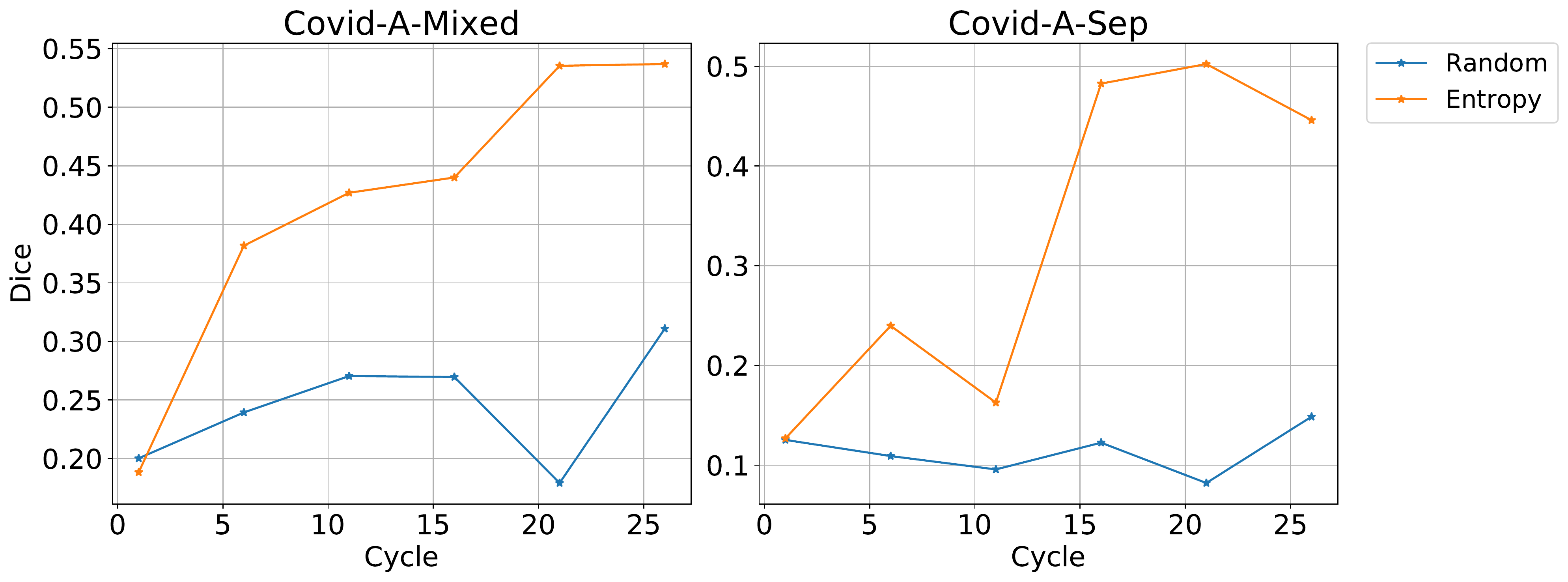}
    \includegraphics[width=0.8\linewidth]{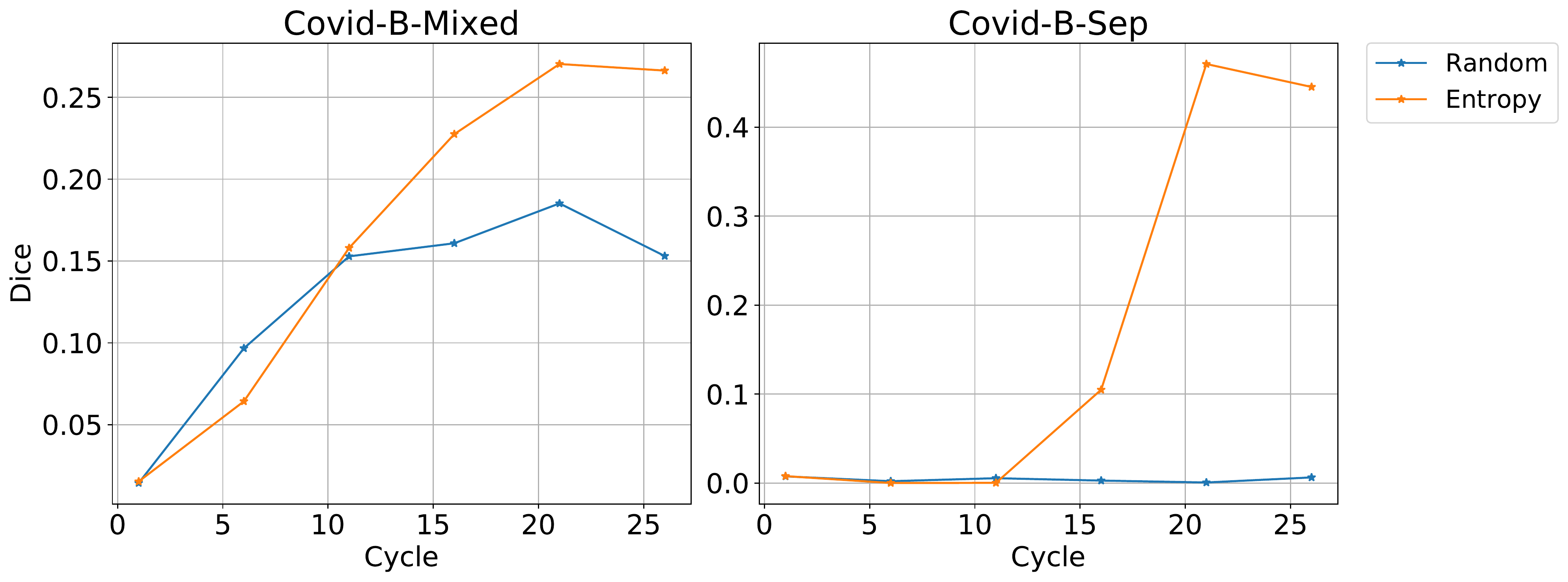}
  \caption{\textbf{Comparison between random and entropy heuristics}. In each cycle, 5 regions were selected for labeling with point-level annotations. Each image is divided into 64 regions. Entropy significantly outperforms random as random tends to select background regions as there is a large imbalance between background and regions with infections.}
  \label{fig:exp1}
\end{figure*}

\subsection{Evaluation Metrics}
As common practice~\cite{shan2020lung}, we evaluate our models using the dice coefficient metric (also known as the F1 Score) for semantic segmentation. Dice is similar to Intersection over Union (IoU)~\cite{everingham2010pascal} but gives more weight to the intersection between the prediction and the ground truth mask, which is computed as $\text{DICE} = \frac{2 * TP}{2 * TP + FP + FN}$,  where TP, FP, and FN is the number of true positive, false positive and false negative pixels across all images in the test set. We also report results with respect to specificity (true negative rate), $Specificity = \frac{TN}{FP + TN}$, which measures the fraction of real negative samples that were predicted correctly.

\begin{figure*}[t]
  \centering
  \includegraphics[width=0.75\linewidth]{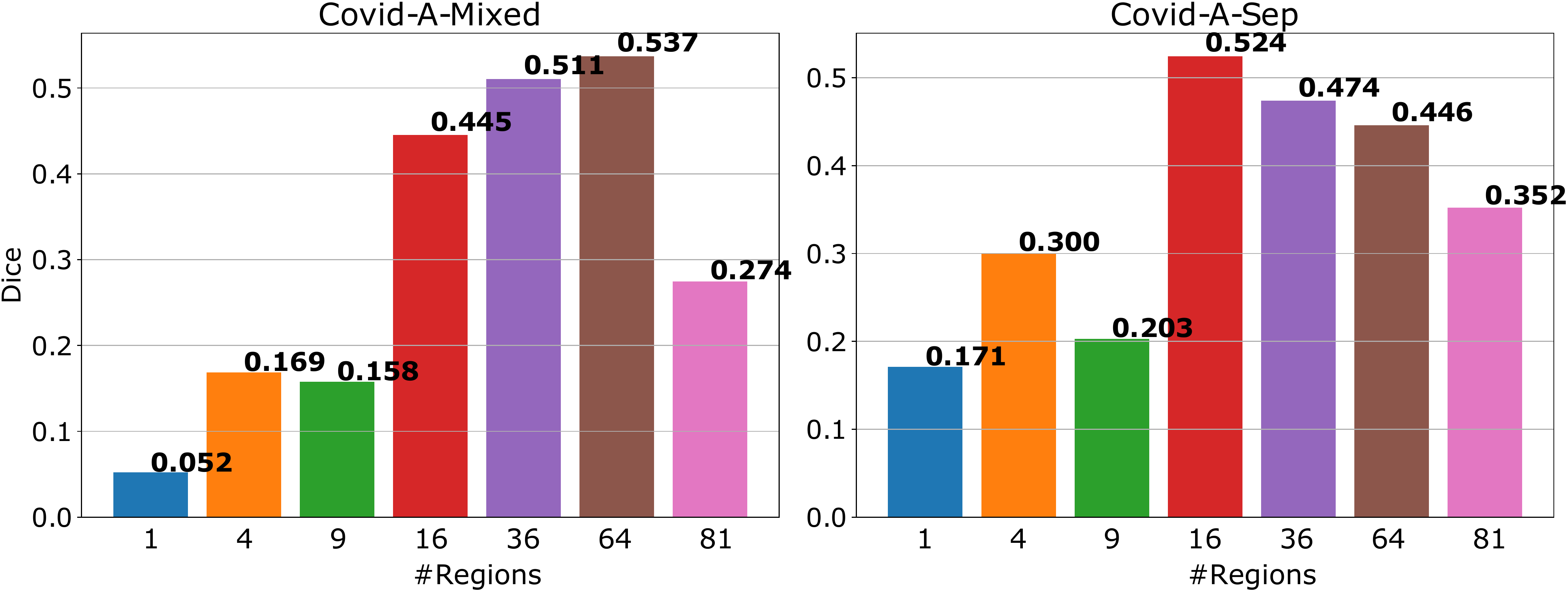}
  \caption{\textbf{Comparison between different region sizes}. For each bar in the plot, the number of regions defines the number of equally-sized non-overlapping rectangles that divide the training images (see rectangle grid in Figure~\ref{fig:model_training}). So higher number of regions means that the regions are of smaller size.}
  \label{fig:exp2}
\end{figure*}

\begin{figure*}[t]
  \centering
  \includegraphics[width=0.85\linewidth]{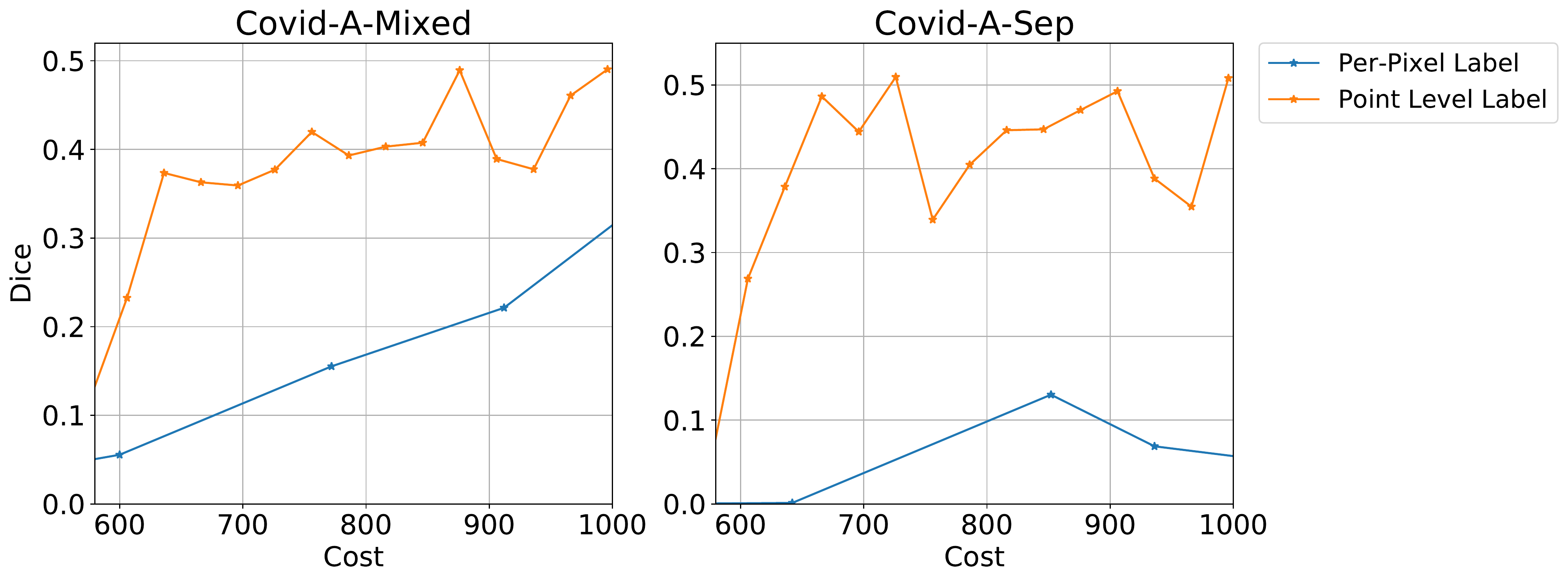}
  \caption{\textbf{Comparison between point-level and per-pixel level supervision based on 64 regions per image}. The cost for a single point annotation is approximated to be 3 seconds~\cite{papadopoulos2017training}. The cost for labeling an infected region is the number of points required to form an approximated polygon around that infection (although in reality it could take more time than that).}
  \label{fig:exp3}
\end{figure*}

\section{Experimental Results}
\label{sec:experiments}


\subsection{Comparing Entropy against Random Heuristic}
For this experiment, a sampled region is labeled in one of two ways depending on whether it contains an infected region. If it has no infected region, then it is labeled with the tag \textit{background}; otherwise the label is a random point annotation on top of an infected region.

The effort required to label a region in either of these two cases is similar. Thus we plot the obtained results regarding the number of labeled regions against the achieved dice score with the trained FCN. We observe in Figure~\ref{fig:exp1} that entropy significantly outperforms the random heuristic for COVID-19-A and COVID-19-B. The reason is that there are many background regions in these two datasets and thus random is more likely to select only background regions, leading to a poor performance. On the other hand, random sampling obtains a good specificity curve ranging between 0.85 and 0.99 as it maintains a high true negative rate. However, false negatives can cost people's lives. Thus it is important to have high recall as well, as achieved with entropy sampling. In other words, as shown in the qualitative results, entropy tends to pick infected regions, as it is where the model is mostly uncertain about. 

For the \textit{separate} splits of COVID-19-A and COVID-19-B, there is a bigger margin between random and entropy and that is because the distribution between the training and testing set is more different than in the \textit{mixed} splits where slices come from the same scans instead. This result suggests a good promise with using region-based active learning with entropy and point-level supervision.

\subsection{Effect of Region Size on Performance}
In this section, we study the impact of the region size on the Dice score performance. The images are divided into $K$ equally-sized non-overlapping regions, so higher number of regions means the regions are of smaller size. In the first cycle, we chose a budget of 192 seconds to label the initial set of images.  For images with 64 regions that corresponds to 3 images ($64 \cdot 3$). Thus, more images are fully labeled for those with smaller number of regions. 

Figure~\ref{fig:exp2} shows that the region size can have a strong impact on the dice performance. Thus, it is important to carefully choose the right region size when using the presented active learning system. For instance, bigger region sizes led to significantly worse performance. The reason is that the annotations might not be placed in the location that could provide the most informative content for the model. Smaller regions focus on where the model is specifically confused at. Further, if the model is confused about the background, selected smaller regions are more likely to contain only background, which provides a strong signal for background.

\subsection{Comparing point-level against per-pixel level supervision}
Here we compare the two labeling schemes: (1) per-pixel label scheme, that is full supervision, and (2) the point-level label scheme. We compute the estimated labeling cost as follows. For the point-level labeler it takes around 3 seconds to make a single point annotation~\citep{papadopoulos2017training}. For the per-pixel labeler we approximate the polygon around the infected mask and use its vertices as the number of points required to annotate that mask. The total effort of labeling the mask is 3 seconds (the cost of a single point label) multiplied by the number of vertices. This cost estimation allows us to compare between the two labeling schemes with respect to the obtained performance.

For the per-pixel level loss function, we combine the weighted cross-entropy and IoU loss as defined in Eq. (3) and (5) from~\citet{wei2019f3net}, respectively. It is an efficient method for ground truth segmentation masks that are imbalanced. Since this loss function requires full supervision, it serves as an upper bound performance in our results.

Figure~\ref{fig:exp3} shows that Point-level labeling achieves superior performance compared to per-pixel labels with lower cost. Each region annotated with per-pixel labels leads to a large increase in labeling cost. Thus, with a fixed annotation budget only few regions can be labeled per-pixel compared with point-level. This result suggests that having more labeled regions with weaker supervision leads to higher overall information content.

\paragraph{Comparison against the upper bound.} We trained the model on the full training set with full supervision on COVID-19-A-Mixed and obtained 84\% Dice score. The total labeling cost is 35328 seconds as the training set consists of 368 slices and  it takes around 96 seconds to label a slice accurately~\cite{ma2020towards}. With our weakly-supervised active learning system and using entropy as our region-selection heuristic, we achieved 76\% dice score (which is around 90\% of the upper bound result) with an effort of 2460 seconds (which is 7\% of the original effort). Figure~\ref{fig:comparisons} shows qualitative results that illustrate that entropy significantly outperforms random with that amount of labeling. For this active learning setup, 5 images were labeled with point-level supervision in the initial cycle. Each image in the training set is divided into 64 regions, leading to an initial effort of $5 \cdot 3 \cdot 64$ seconds. In each cycle, for 100 cycles, we labeled 5 regions with point-level annotations, leading to a total cost of $5 \cdot 3 \cdot 64 + 100 \cdot 5 \cdot 3 = 2460$. This result suggests that we can achieve a strong performance with very low human effort. 

\section{Conclusion}
We have proposed a weakly supervised region-based active learning setup for cost-efficient labeling of COVID19 infections in CT scans. This framework combines two ideas for reducing labeling effort. The first idea is to use a region-based active learning approach which, different from conventional active learning, presents the annotator with regions of the image for instead of the whole image. Using entropy-based MCMC, this scheme encourages labeling highly informative regions that maximize the model's validation accuracy. The second idea is to use point-level supervision which is much cheaper to acquire than per-pixel labels. Since this labeling scheme requires the annotator to label each infected region with a single click, it only requires 3 seconds per infected region. Our results show that entropy-based heuristics outperform random selection of regions with respect to the dice score and labeling effort. Moreover, we show that region-based annotation outperforms whole image labeling in terms of cost efficiency. As a result, our system reaches around 90\% of the dice score of the same model trained on the whole training set with only 7\% of the effort. For future work, it would be interesting to investigate other forms of weak supervision and other forms of regions that are not rectangles. Such regions could include super pixels or selective search proposals.

{\small
\bibliographystyle{abbrvnat}
\bibliography{egbib}
}

\end{document}

%% file: math_commands.tex
\usepackage{amsmath,amsfonts,bm}

\def\eqref#1{equation~\ref{#1}}

\def\1{\bm{1}}

\DeclareMathAlphabet{\mathsfit}{\encodingdefault}{\sfdefault}{m}{sl}
\SetMathAlphabet{\mathsfit}{bold}{\encodingdefault}{\sfdefault}{bx}{n}

